\documentclass[aps,preprint,12pt,showpacs,groupedaddress,onecolumn]{revtex4}
\usepackage{amsfonts}
\usepackage{amsmath}
\usepackage{amssymb}
\usepackage{graphicx}

\begin{document}

\title{Ballistic Thermal Transistor of Dielectric Four-terminal
Nanostructures}
\author{Ping Yang$^{1,}$}
\email{yangpingg@gmail.com}
\author{Bambi Hu$^{1,2}$}
\affiliation{$^{1}$Department of Physics, Centre for Nonlinear Studies, and The
Beijing-Hong Kong-Singapore Joint Centre for Nonlinear and Complex Systems
(Hong Kong), Hong Kong Baptist University, Kowloon Tong, Hong Kong, China
\linebreak$^{2}$Department of Physics, University of Houston, Houston, Texas
77204-5005, USA}

\begin{abstract}
We report a theoretical model for a thermal transistor in dielectric
four-terminal nanostructures based on mesoscopic ballistic phonon
transport, in which a steady thermal flow condition of system is
obtained to set up the temperature field effect of gate. In the
environment, thermal flow shows the transisting behaviors at low
temperatures: saturation, asymmetry, and rectification. The
phenomena can be explained reasonably by the nonlinear variation of
the temperature dependence of propagating phonon modes in terminals.
The results suggest the possibility of the novel nano-thermal
transistor fabrication.
\end{abstract}

\maketitle

Rapid advances in nanometer-sized techniques have made possible the
miniaturization and integration of electronic devices. Usually these device
properties cannot be described by classical transport theory, since their
characteristic sizes are small in comparison with the elastic mean free path
between scattering events, particularly at low temperature. In this
situation, the wave nature of the electrons needs to be taken explicitly
into account, and electrons do not propagate diffusively instead
ballistically. Based on the ballistic and phase coherent electron transport,
a variety of interesting devices, e.g., the quantum stub transistor, the
nanotube transistor, and the multi-terminal junctions, have been proposed%
\cite{ref1}.

As the counterpart of electron transport, mesoscopic phonon
transport has been paid much attention recently. The new phenomenon
of the universal quantum of thermal conductance at low temperature
was predicted theoretically in mesoscopic dielectric
systems\cite{ref2}, and verified experimentally\cite{ref3}. Some
investigations of thermal transport properties have been done
sequentially at the variety of
nanostructures\cite{ref4,ref5,ref6,ref7,ref8,ref9}. However the
thermal transistor actions in mesoscopic dielectric systems, based
on quantum interference effect, have to our knowledge not been
studied either theoretically or experimentally. It should have
different phenomena and mechanism from that of the work based on
classical theory\cite{ref10}.

In this paper, we demonstrate a theoretical model of a thermal
transistor in dielectric four-terminal nanostructures (DFNSs) based
on the ballistic nature of phonon transport at low temperatures, in
which a steady thermal flow condition of system is obtained to set
up the temperature field effect of gate. Therefore the DFNSs can
work as a ballistic thermal transistor, with saturation of thermal
flow from source to drain for the gate temperature at large
temperature difference between the source and drain. For the
asymmetric structure, thermal asymmetry and rectification are showed
under changes in the temperature difference. A reasonable
explanation of the phenomena is given by the nonlinear variation of
the temperature dependence of propagating phonon modes in terminals.
The possibility of the experimental observation of these novel
phenomena is also discussed.

\begin{figure}[tbp]
\begin{center}
\includegraphics[
height=3.301in,
width=4.3267in
]{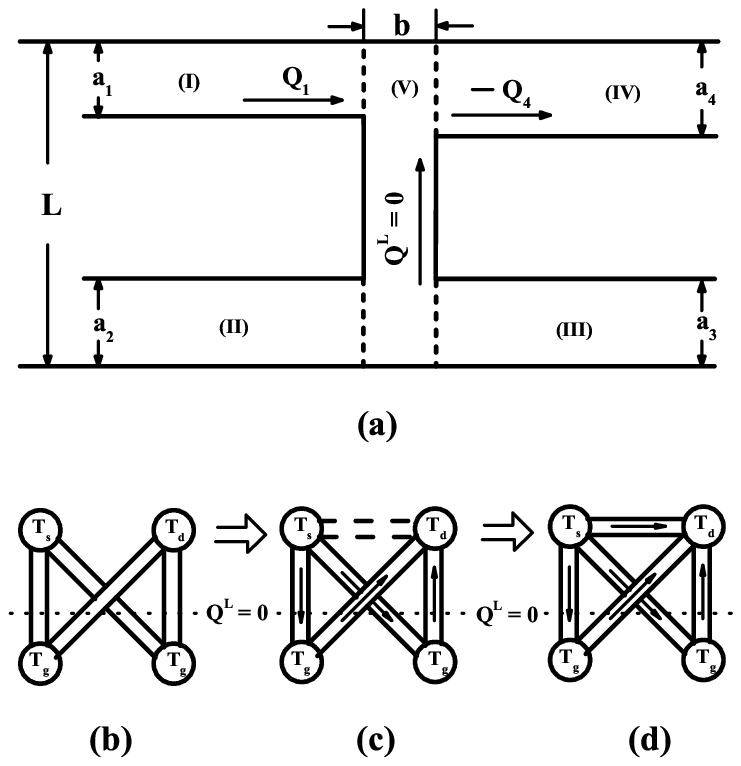}
\end{center}
\caption{Schematic diagrams for the research system and the working
illustration of SGD. Arrow lines represent the thermal flows. (a) A
four-terminal structure. (b) SGD at equilibrium:
$T_{s}=T_{g}=T_{d}$. (c) SGD at steady state: $T_{s}>T_{g}>T_{d}$; a
propagating channel (dash lines) is set up between the source and
drain. (d) Thermal flow profile at steady state. $Q_{1}$ is the net
thermal flow in terminal $1$, and $Q^{L}$ is the net thermal flow
across the level line (dot line) or (arrow line) in the vertical
direction of region $V$ and always equals zero at steady state of
system.} \label{FIG1}
\end{figure}

A two-dimensional DFNS is illustrated in Fig.\ref{FIG1}(a). Four
parallel terminal wires of width $a_{i}$ are directly coupled to a
central ballistic window region $V$ with width $b$ and height $L$
(scattering region). Other ends of four-terminal wires are connected
to thermal reservoirs at equilibrium with temperatures $T_{i}$,
$i=1,2,3,4,$ respectively. Supposed that there is no phonon
scattering inside the terminal wires and the central window, and a
perfect connection between terminal wires and reservoirs, phonon
scattering is solely decided by the geometrical features and happens
only at interfaces between the terminal wires and the central
window. The net thermal flow $Q_{i}$ in
terminal wire $i$ is expressed as\cite{ref6}%
\begin{equation}
Q_{i}=\sum_{j(j\neq
i)}Q_{ij}=\sum\limits_{j(j\not=i)}\sum\limits_{m,n}\int_{max(\omega
_{im},\omega _{jn})}^{\infty }\frac{d\omega }{2\pi }\hbar \omega \lbrack
n_{i}(\omega ,T_{i})-n_{j}(\omega ,T_{j})]\widetilde{T}_{ji,nm}(\omega )\ \ ,
\label{a1}
\end{equation}%
where $n_{i}(\omega ,T_{i})=[exp(\hbar \omega /k_{B}T_{i})-1]^{-1}$ is the
Bose-Einstein distribution function of the phonons with temperature $T_{i}$
in the $i$-th reservoir, $k_{B}$ is the Boltzmann constant, and $\hbar $ is
Planck's constant. $\omega _{im}=\frac{\pi vm}{a_{i}}$ is the cutoff
frequency of mode $m$ in terminal wire $i$, and $v$ ($=5000$ $m\cdot
sec^{-1} $) is the group velocity. $\widetilde{T}_{ji,nm}(\omega )$ is the
phonon transmission coefficient that an incident phonon with energy $\hbar
\omega $ from terminal $i$ at phonon mode $m$ is scattered to terminal $j$
at mode $n$, with the property: $\widetilde{T}_{ji,nm}(\omega )=\widetilde{T}%
_{ij,mn}(\omega )$. For the present DFNSs, transmission coefficient
can be obtained by the mode matching method, the same approach as
Ref.\cite{ref6,ref7}. Thermal flow $Q_{ij}$ is decided by two
thermal reservoirs $i$ and $j$, and has the property:
$Q_{ij}=-Q_{ji}$.

In the paper, we assume that the dielectric four-terminal system has
reached the steady state. The steady thermal flow across central
window from terminal 1 to terminal 4 has the definition:
$Q_{1}=-Q_{4}$, where the plus$/$minus sign represents separately
the thermal flow into$/$out of system. Meanwhile temperatures of
four thermal reservoirs are assumed to have the sequence:
$T_{1}\geqslant (T_{2},T_{3})\geqslant T_{4}$. In terms of
Eq.(\ref{a1}) and the steady flow
definition, we readily obtain the steady thermal flow condition of system:%
\begin{equation}
Q_{12}+Q_{13}+Q_{42}+Q_{43}=0\text{ \ }.  \label{a2}
\end{equation}%
If adding the ($Q_{23}+Q_{32}$) into Eq.(\ref{a2}), we also get
$Q_{2}=-Q_{3} $, the steady flow between terminals 2 and 3.
Obviously, the temperatures of four thermal reservoirs, which are
satisfied with Eq.(\ref{a2}), can make the DFNS at steady state.
Note that the summation of four terms in the left hand side of
Eq.(\ref{a2}) is just the thermal flow $Q^{L}$ along the vertical
direction of central window (region $V$), which equals zero in terms
of Eq.(\ref{a2}), (see in Fig.\ref{FIG1}(a)). This means that the
thermal flow across terminals $1$ and $4$ isolates from that across
terminals $2$ and $3$ at steady state of system. With the help of
Eq.(\ref{a1}), equation (\ref{a2}) transforms to
\begin{equation}
\left[ Q_{12}^{s}(T_{1})+Q_{13}^{s}(T_{1})\right] +\left[
Q_{42}^{s}(T_{4})+Q_{43}^{s}(T_{4})\right] =\left[
Q_{21}^{s}(T_{2})+Q_{24}^{s}(T_{2})\right] +\left[
Q_{31}^{s}(T_{3})+Q_{34}^{s}(T_{3})\right] \text{ },  \label{a3}
\end{equation}%
where $Q_{ij}^{s}(T_{i})$ is the thermal flow that is produced by single
thermal reservoir with temperature $T_{i}$ and that flows directly from
terminal $i$ into terminal $j$; $Q_{ij}^{s}(T_{i})=\sum_{m}\int_{\omega
_{im}}^{\infty }\frac{d\omega }{2\pi }\hbar \omega n_{i}(\omega ,T_{i})%
\widetilde{T}_{ij,m}(\omega ),$ $\widetilde{T}_{ij,m}(\omega
)=\sum_{n}\theta (\omega -\omega _{jn})\widetilde{T}_{ij,mn}(\omega
)$. In this way, the steady flow condition of system of
Eq.(\ref{a2}) can be expressed by four single thermal reservoirs. We
take reservoir $1
$ for the source with $T_{s}=T_{1}$, reservoir $4$ for the drain with $%
T_{d}=T_{4}$, and reservoirs $2$ and $3$ for the gates with $%
T_{g}=T_{2}=T_{3}$ respectively. The simplified form of
Eq.(\ref{a3}) is written as
\begin{equation}
E_{1}^{s}(T_{s})+E_{4}^{s}(T_{d})=E_{2}^{s}(T_{g})+E_{3}^{s}(T_{g})\text{ ,}
\label{a4}
\end{equation}%
where $E_{1}^{s}(T_{s})=Q_{12}^{s}(T_{s})+Q_{13}^{s}(T_{s})$, $%
E_{4}^{s}(T_{d})=Q_{42}^{s}(T_{d})+Q_{43}^{s}(T_{d})$, $%
E_{2}^{s}(T_{g})=Q_{21}^{s}(T_{g})+Q_{24}^{s}(T_{g})$, and $%
E_{3}^{s}(T_{g})=Q_{31}^{s}(T_{g})+Q_{34}^{s}(T_{g})$. Equation
(\ref{a4}) sets up the source-gate-drain (SGD) relation, in which
the gates connect the source and drain respectively and work as
controlling ends. For a fixed gate temperature $T_{g}$, a pair of
corresponding temperatures $T_{s}$ and $T_{d}$ can be obtained from
Eq.(\ref{a4}). Each set of temperatures including the $T_{g}$ and a
pair of $T_{s}$ and $T_{d}$ can make sure the system at steady state
and be used to calculate the steady thermal flow $Q_{sd}$ from
source to drain in terms of Eq.(\ref{a1}); $Q_{sd}=Q_{1}$ in
terminal $1$ and $Q_{sd}=-Q_{4}$ in terminal $4$. As stated above,
we can get all values of the steady thermal flow $Q_{sd}$
corresponding to different temperature pairs of $T_{s}$ and $T_{d}$
that set up the temperature difference between source and drain,
$i.e.$, $T_{sd}=T_{s}-T_{d}$, for the fixed $T_{g}$ in the DFNSs.
The schematic diagrams of working procedure for $Q_{sd}$ is shown in
Fig.\ref{FIG1}(b)$-$(d). Furthermore, as $Q^{L}=0$, the influence of
the gate temperature $T_{g}$ on the steady thermal flow $Q_{sd}$
should be attributed to the temperature field effect (TFE). Note
that all temperatures in the paper always mean those of thermal
reservoirs.

\begin{figure}[ptb]
\begin{center}
\includegraphics[
height=3.7619in,
width=5.2771in
]{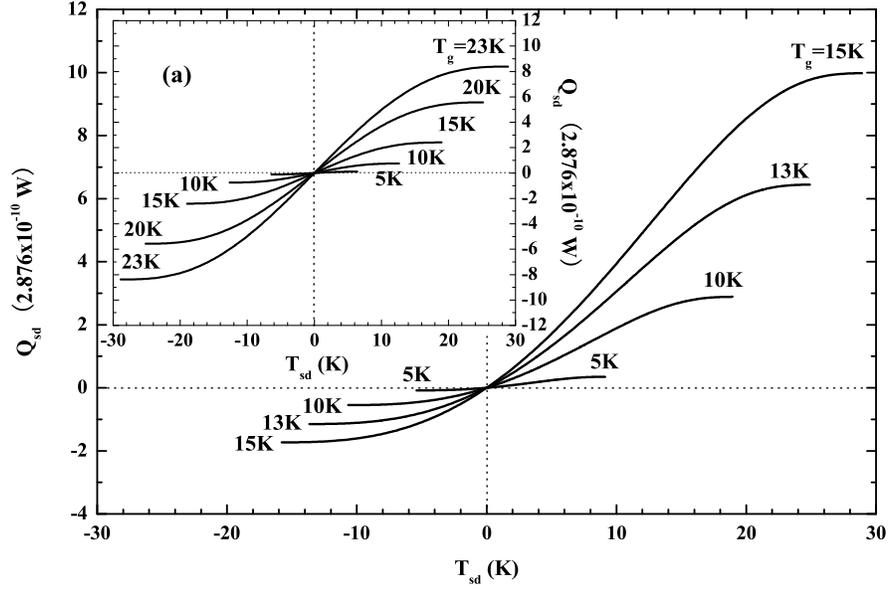}
\end{center}
\caption{Thermal flow $Q_{sd}$ vs temperature difference $T_{sd}$
for the gate temperature $T_{g}$ in the DFNS, the same structure as
shown in Fig.\protect\ref{FIG1}(a). [Main figure] Asymmetric
structure with $a_{4}=40nm$. [Inset (a)] Symmetric structure with
$a_{4}=20nm$. Other parameters: $a_{1}=a_{2}=a_{3}=20nm$, $L=100nm$,
$b=20nm$.} \label{FIG2}
\end{figure}

\begin{figure}[ptb]
\begin{center}
\includegraphics[
height=3.0217in, width=4.2134in ]
{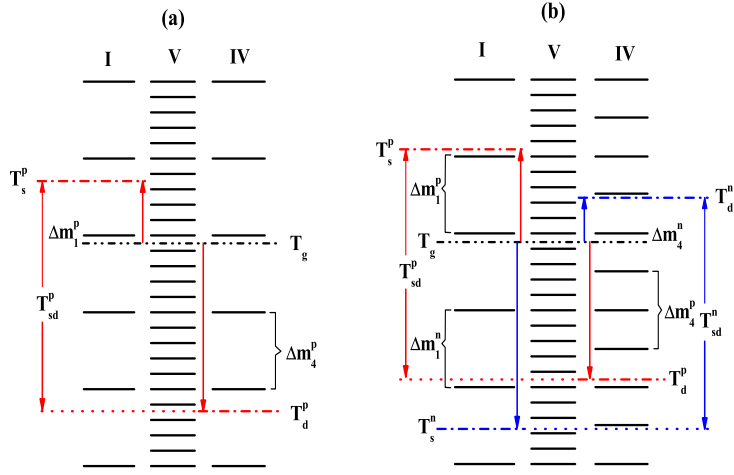}
\end{center}
\caption{Schematic diagrams for the cutoff frequencies of discrete
transverse phonon modes or threshold energy (straight lines) for
terminal $I$, region $V$, and terminal $IV$. Dash dot dot line:
the gate temperature $T_{g}$. Dash dot lines: temperatures $T_{s}$ and $%
T_{d} $. Arrow lines: the change of temperatures relative to
$T_{g}$. The superscript $p$ represents the positive $T_{sd}$
($T_{s}>T_{d}$); $n$ for the negative $T_{sd}$ ($T_{s}<T_{d}$).
$\Delta m_{1}^{p}$ is modes in terminal $1$ excited by $T_{s}^{p}$.
(a) Symmetric structure. (b) Asymmetric structure for
$T_{sd}^{p}=T_{sd}^{n}$.}
\label{FIG3}
\end{figure}

\begin{figure}[ptb]
\begin{center}
\includegraphics[
height=3.4973in, width=4.689in ] {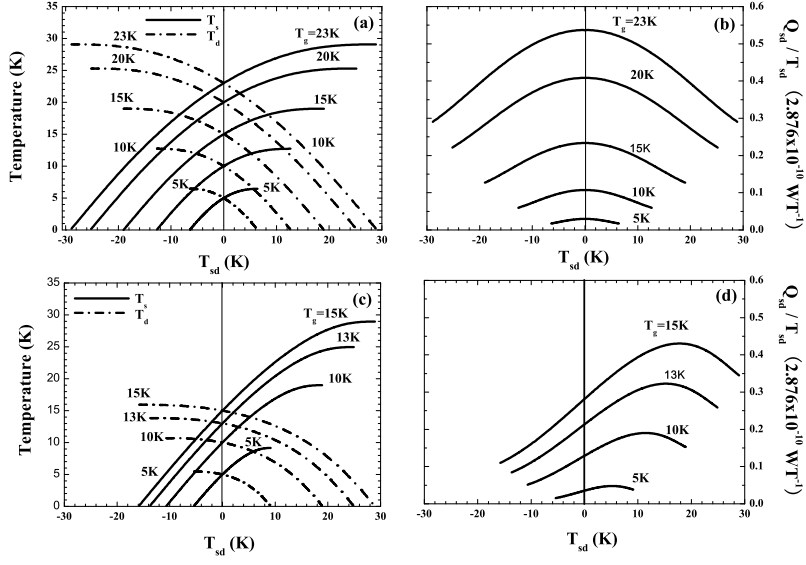}
\end{center}
\caption{At steady state of system, for the fixed gate temperatures
$T_{g}$s: [(a) and (c)] Temperatures $T_{s}$ in source and $T_{d}$
in drain vs temperature difference $T_{sd}$ between the source and
drain, respectively
for the symmetric and asymmetric cases. [(b) and (d)] Thermal conductance $%
Q_{sd}/T_{sd}$ vs $T_{sd}$ respectively for the symmetric and
asymmetric cases. }
\label{FIG4}
\end{figure}

For the symmetric DFNS, the inset(a) of Fig. \ref{FIG2} shows the
calculated thermal flow $Q_{sd}$ as a function of the temperature
difference $T_{sd}=T_{s}-T_{d}$ for five different values of
$T_{g}$. As seen in the figure, $Q_{sd}$ first increases, and then
tends to saturation with $T_{sd}\ $ increasing for the five
$T_{g}$s. In the following, we will study the phenomenon based on
mesoscopic ballistic transport theory. Figure \ref{FIG3} illustrates
the cutoff frequencies of the discrete transverse phonon modes
(threshold energy) respectively in terminal $1$, scattering region
$V$, and terminal $4$, where the superscript $p$ and $n$ represent
separately the positive $T_{sd}$ at $T_{s}>T_{d}$ and the negative
$T_{sd}$ at $T_{s}<T_{d}$. For the symmetric DFNS with
$a_{1}=a_{2}=a_{3}=a_{4}=20nm$, as shown in Fig.\ref{FIG3}(a), the
mode spacing $\Delta_{i} =\omega_{i(m+1)}-\omega_{im}=\frac{\pi
v}{a_{i}}$ in terminal $1$
equals that of terminal $4$. At initial state with $%
T_{s}^{p}=T_{d}^{p}=T_{g}$, the system is at equilibrium ($T_{sd}^{p}=0$, $%
Q_{sd}=0)$. Then the enhance of $T_{s}^{p}$ relative to the fixed
$T_{g}$ excites the additional phonon modes $\Delta m_{1}^{p}$ in
terminal $1$ and makes energy into terminals $2$ and $3$ from all
modes $(m_{1}^{eq}+\Delta m_{1}^{p})$ in terminal $1$, in terms of
$\frac{\pi v\Delta m_{1}^{p}}{a_{1}}$ $\backsim k_{B}\left\vert
T_{s}^{p}-T_{g}\right\vert /\hbar $, where the superscript $eq$
represents the equilibrium state. Since $Q_{ij}^{s}(T_{i})$ in
Eq.(\ref{a3}) is monotonously increasing function of temperature
$T_{i}$\cite{ref6,ref7}, $E_{1}^{s}(T_{s})$ in Eq.(\ref{a4})
increases correspondingly. Meanwhile, to balance the both sides in
Eq.(\ref{a4}), $E_{4}^{s}(T_{d})$ must have equal amount decreased
because $E_{2}^{s}(T_{g})$ and $E_{3}^{s}(T_{g})$ in Eq.(\ref{a4})
keep constant. This means that the modes $m_{4}^{eq}$ in terminal
$4$ have $\Delta m_{4}^{p}$ decrease and the corresponding
temperature $T_{d}^{p}$ drops below $T_{g}$ as shown in
Fig.\ref{FIG3}(a). Since $\Delta m_{1}^{p}$ excited by $T_{s}^{p}$
at $T_{s}^{p}>T_{g}$ are higher energy modes, the decreased energy
of $\Delta m_{4}^{p}$ which equals $\Delta m_{1}^{p}$ is not
sufficient to cancel out the additional energy produced by
$T_{s}^{p}$. Therefore $\Delta m_{4}^{p}$ in
terminal $4$ must be more than $\Delta m_{1}^{p}$ in terminal $1$, $i.e.$, $%
\Delta m_{4}^{p}>$ $\Delta m_{1}^{p}$, in order to make the
increased energy from ($m_{1}^{eq}+$ $\Delta m_{1}^{p}$) in terminal
$1$ and the decreased energy from $\Delta m_{4}^{p}$ in terminal $4$
equal. This means that there is an unequal amount change of $\Delta
m_{1}^{p}$ and $\Delta
m_{4}^{p}$ for the equal amount of $E_{1}^{s}(T_{s})$ increase and $%
E_{4}^{s}(T_{d})$ decrease in Eq.(\ref{a4}). With $T_{sd}^{p}$
increasing further, as $\Delta m_{i}^{p}\backsim \left\vert
T_{i}^{p}-T_{g}\right\vert $, the ratio of $\Delta m_{1}^{p}$ to
$\Delta m_{4}^{p}$ is the nonlinear monotonously decreasing;
therefore $T_{s}^{p}$ shows the nonlinear increasing behavior,
meanwhile $T_{d}^{p}$ has the nonlinear decreasing behavior as shown
in Fig.\ref{FIG4}(a), in order to keep the system at steady state.
Obviously the modes $(\Delta m_{1}^{p}+\Delta m_{4}^{p})$
corresponding to $(T_{s}^{p}-T_{d}^{p})$ become more with
$T_{sd}^{p}$ increasing, but the ratio of $\Delta m_{1}^{p}$ to
$(\Delta m_{1}^{p}+\Delta m_{4}^{p})$ goes down. This means that $%
T_{s}^{p}$ has less contribution than $T_{d}^{p}$ for $T_{sd}^{p}$
increasing in terms of $\Delta m_{i}^{p}\backsim \left\vert
T_{i}^{p}-T_{g}\right\vert $. On the other hand, as the propagating
channel between the source and drain consists of the propagating
phonon
modes ($m_{1}^{eq}+\Delta m_{1}^{p}$) in terminal $1$ and ($%
m_{4}^{eq}-\Delta m_{4}^{p}$) in terminal $4$, $Q_{sd}$
increases with $T_{sd}^{p}$; but the increment rate of thermal conductance $%
Q_{sd}/T_{sd}^{p}$ reduces with $T_{sd}^{p}$ increasing, $i.e.$, $%
Q_{sd}/T_{sd}^{p}>0$ and
$\frac{d}{dT_{sd}^{p}}(Q_{sd}/T_{sd}^{p})<0$, (see in
Fig.\ref{FIG4}(b)). Consequently, under the influence of $T_{g}$,
the nonlinear variation of the modes $\Delta m_{1}^{p}$ and $\Delta
m_{4}^{p}$
in the modes $(\Delta m_{1}^{p}+\Delta m_{4}^{p})$ that corresponds to $%
T_{sd}^{p}$ leads to the saturation behavior of thermal flow
$Q_{sd}$ in the DFNS. Furthermore, as the mode spacings in terminals
$1$ and $4$ are identical in this symmetric structure, $Q_{sd}$
presents symmetry in the reversal of $T_{sd}^{p}$, $i.e.$,
$T_{d}^{n}>T_{s}^{n}$. Additionally, $Q_{2}$($=-Q_{3}$) is two
orders of magnitude lower than $Q_{sd}$.

In order to investigate the asymmetric effect, we change the DFNS
from the symmetric structure to the asymmetric with
$a_{1}=a_{2}=a_{3}=20nm,a_{4}=40nm$. As shown in Fig.\ref{FIG2},
there is asymmetry in the $Q_{sd}-$ $T_{sd}$ behavior with thermal
flow saturation for the positive and negative $T_{sd}$ for four
values of $T_{g}$. (i) Thermal saturation discussion: for the
asymmetric DFNS, the mode spacing $\Delta _{1}=\frac{\pi v}{a_{1}}$
in terminal $1$ is twice $\Delta _{4}$ in terminal $4$; there is a
higher density of transverse phonon modes in terminal $4$ as shown
in Fig.\ref{FIG3}(b). When $T_{s}^{p}$ increases over $T_{g}$,
thermal energy input from the modes ($m_{1}^{eq}+\Delta m_{1}^{p}$)
in terminal $1$ is canceled out by the decreased modes $\Delta
m_{4}^{p}\ $along with $T_{d}^{p}$ lowing, to satisfy the SGD
relation, (see in Fig.\ref{FIG3}(b)and Fig.\ref{FIG4}(c)). However,
because there is higher density of phonon modes in terminal $4$ than
that in the symmetric situation, at initial stage, the small change
of $T_{d}^{p}$ could make sufficient modes $\Delta m_{4}^{p}$
decreased to cancel out the energy input from $T_{s}^{p}$ enhance;
the ratio of $\Delta m_{1}^{p}$ to $(\Delta m_{1}^{p}+\Delta
m_{4}^{p})$ drops less with $T_{sd}^{p}$ increasing, in comparison
with the symmetric situation. Therefore $Q_{sd}/T_{sd}^{p}$
increases with $T_{sd}^{p}$ increasing, $i.e.$,
$Q_{sd}/T_{sd}^{p}>0$ and
$\frac{d}{dT_{sd}^{p}}(Q_{sd}/T_{sd}^{p})>0$, (see in
Fig.\ref{FIG4}(d)). With $T_{s}^{p}$ increasing further, more
$\Delta m_{1}^{p}$ with higher energy in terminal $1$ are excited
and more energy is input into system; therefore $\Delta m_{4}^{p}$
in terminal $4$ reduce more and $T_{d}^{p}$ correspondingly goes
much down to balance the both sides in Eq.(\ref{a4}). The ratio of
$\Delta m_{1}^{p}$ to $(\Delta m_{1}^{p}+\Delta m_{4}^{p})$
decreases remarkably with $T_{sd}^{p}$, similar to the symmetric
situation. Thus $Q_{sd}/T_{sd}^{p}$ changes from increase to
decrease with $T_{sd}^{p}$, as shown in Fig.\ref{FIG4}(d);
$\frac{d}{dT_{sd}^{p}}(Q_{sd}/T_{sd}^{p})>0$ at beginning, and then
$\frac{d}{dT_{sd}^{p}}(Q_{sd}/T_{sd}^{p})<0$ after $T_{sd}^{p}$
passing a critical value. This means that $Q_{sd}-T_{sd}^{p}$ curve
increases monotonously with the concave shape at beginning, and then
keeps going with the convex shape after passing a turning point, and
finally tends to saturation, (see in Fig.\ref{FIG2}). (ii) Thermal
asymmetry and rectification discussion: in the reverse case with
$T_{d}^{n}>$ $T_{s}^{n}$, due to the higher density of phonon modes
in terminal $4$, $T_{d}^{n}$ enhance over $T_{g}$ could easily
excite more $\Delta m_{4}^{n}$ in terminal $4$ and makes more energy
into system. This leads that $\Delta m_{1}^{n}$ in terminal $1$ must
decrease much, due to the larger mode spacing $\Delta =\frac{\pi
v}{a_{1}}$ in terminal $1$. Thus $T_{s}^{n}$ corresponding to
$\Delta m_{1}^{n}$ decreases more, in terms of $\frac{\pi v\Delta
m_{1}^{n}}{a_{1}}$ $\backsim k_{B}\left\vert
T_{s}^{n}-T_{g}\right\vert /\hbar $, (see in Fig.\ref{FIG3}(b)). The
ratio of $\Delta m_{4}^{n}$ to $(\Delta m_{1}^{n}+\Delta m_{4}^{n})$
decreases rapidly with $T_{sd}^{n}$, so that $Q_{sd}$ tends more
quickly to saturation in the reverse case as shown in
Fig.\ref{FIG2}. Furthermore for the fixed $T_{sd}$: $\left\vert
T_{s}-T_{d}\right\vert
=(T_{s}^{p}-T_{d}^{p})=(T_{d}^{n}-T_{s}^{n})=$ constant,the $\Delta
m_{1}^{p}$ with higher energy for the positive case
$(T_{s}^{p}>T_{d}^{p})$ is more than the $\Delta m_{4}^{n}$ for the
reverse
case $(T_{s}^{n}<T_{d}^{n})$; therefore we have $T_{s}^{p}>$ $T_{d}^{n}$ and $%
T_{d}^{p}>$ $T_{s}^{n}$, (see in Fig.\ref{FIG3}(b)). Thus, in the positive case, $%
T_{s}^{p}$ and $T_{d}^{p}$ locate at higher temperature region to
make more modes with higher energy participated in thermal
transport; for the reverse case, $T_{s}^{n}$ and $T_{d}^{n}$ is
situated at lower temperature region to excite less modes with
higher energy for thermal transport, (see in Fig.\ref{FIG4}(c) and
Fig.\ref{FIG3}(b)). Consequently the $Q_{sd}-T_{sd}$ characteristics
display a distinct asymmetry and rectifying behavior in the
asymmetric structure as shown in Fig.\ref{FIG2}. In the above
discussion, the main effect of $T_{g}$ is to set up a reference
temperature which makes the nonlinear variation of $\Delta m_{1}$
and $\Delta m_{4}$ possible.

In view of some successful thermal measurements in
nanoscale\cite{ref3,ref11,ref12}, it is possible to realize the
relevant experiments for the novel phenomena in the DFNSs. Authors
think that the steady thermal flow in the ballistic system is a
necessary factor to observe these thermal transisting behaviors as
discussed in the paper.

In summary, thermal transistor has been designed theoretically in
the DFNSs. When satisfied with the steady thermal flow condition of
system, saturation and asymmetry and rectification of thermal flow
are exhibited separately in the symmetric and asymmetric DFNSs. The
relevant physical mechanism is elucidated by the nonlinear variation
of temperature dependence of phonon modes in terminals. The
mechanism is also helpful to investigation of the negative
differential thermal resistance and thermal diode design in
nanoscale systems.

This work was supported through grants by the Hong Kong Research
Grants Council (RGC) and the Hong Kong Baptist University Faculty
Research Grant (FRG).


\newpage

\end{document}